**Title**: An ARIMA model to forecast the spread and the final size of COVID-2019 epidemic in Italy.
**Author**. Gaetano Perone. **Affiliation.** University of Bergamo.

**Abstract**

Coronavirus disease (COVID-2019) is a severe ongoing novel pandemic that is spreading quickly across the world. Italy, that is widely considered one of the main epicenters of the pandemic, has registered the highest COVID-2019 death rates and death toll in the world, to the present day. In this article I estimate an autoregressive integrated moving average (ARIMA) model to forecast the epidemic trend over the period after April 4, 2020, by using the Italian epidemiological data at national and regional level. The data refer to the number of daily confirmed cases officially registered by the Italian Ministry of Health (www.salute.gov.it) for the period February 20 to April 4, 2020. The main advantage of this model is that it is easy to manage and fit. Moreover, it may give a first understanding of the basic trends, by suggesting the hypothetic epidemic's inflection point and final size.

**Keywords**: COVID-2019; infection disease; pandemic; time series; ARIMA model; forecasting models.

**JEL CODES**: C22; C53; I18

**Highlights**:
- ARIMA models allow in an easy way to investigate COVID-2019 trend.

- All governmental institutions, especially in public health, may benefit from these data.

- These data may be used to monitor the epidemic and to better allocate the resources.

- Further useful and more precise forecasting may be provided by updating these data or applying the model to other regions and countries.

## 1. Introduction

Coronavirus disease (COVID-2019) is a severe ongoing novel pandemic that has emerged in Hubei, a central province of China, in December 2019. In few months, it has spread quickly across the world, and at the time of writing has affected more than 200 countries and has caused about tens of thousands of deaths. The most affected country are China, France, Germany, Italy, Spain, and USA. Italy is considered one of the main epicentres of the pandemic due to its pretty high death rates (12.33%) and death toll (15,362),[1] and it represents the nucleus of this short paper.

When an epidemic occurs, one of the crucial questions is to determine its evolution and inflection point. So, the main aim of this paper is to provide a short-term forecast of the COVID-2019 diffusion in Italy, by using an autoregressive integrated moving average (ARIMA) model on national and selected regional data, over the period after April 4, 2020. The paper is organized as follows. In section 2, I will provide an explanation of the data used in the econometric analysis. In section 3 I will discuss the empirical strategy and the main results of ARIMA models. Finally, in section 4 I will stress the possible meaning of the results.

---

[1] On April 4, 2020, Italy had the world's highest death rates and death toll due to COVID-2019.

## 2. Data description

The data used in this analysis refer to the number of new daily COVID-2019 confirmed cases from February 20, 2020 to April 4, 2020, and are extracted from the official website of the Italian Ministry of Health (www.salute.gov.it). They include the overall national trend and five selected Italian regions: Emilia Romagna, Lombardy, Marche, Tuscany, and Veneto. Marche and Tuscany belong to the centre of Italy, while Emilia Romagna, Lombardy, and Veneto belong to the north of Italy. These regions have been chosen because of their centrality in the Italian outbreak; in fact, they are characterized by the highest number of COVID-2019 confirmed cases on April 4, 2020. Lombardy is the country's leading region, with a mortality rate of 17,62% and 49,118 confirmed cases, the 39.41% of the overall Italian cases, followed by Emilia Romagna (16,540 cases), Veneto (10,824 cases), Tuscany (5,671 cases), and Marche (4,341).[2]
About 79.4% of COVID-2019 cases are concentrated in the north of the country. This clearly shows that COVID-2019 has especially affected the north of the country.

Figure 1. New daily COVID-2019 confirmed cases in the Italian regions since the start of the epidemic.

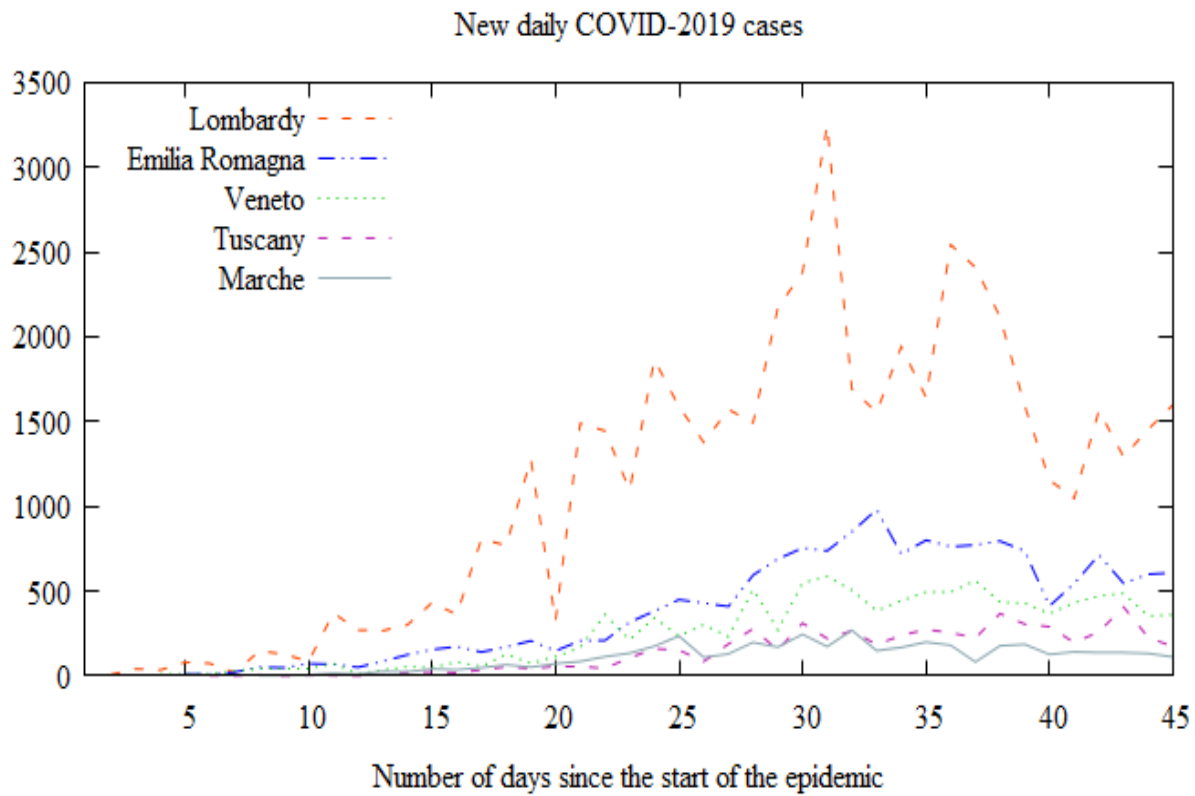

Notes: author's elaboration on Italian Ministry of Health data (www.salute.gov.it).[3]

The descriptive analysis of the overall and regional data shows that the new daily COVID-2019 confirmed cases have increased approximately until the 37th-38th day since the start of the epidemic. Then, they have showed a decreasing trend, by suggesting a possible epidemic slowdown (Figure 1). I will try to analyse this trend in the next section.

---

[2] It's necessary to stress that Piemonte has registered 11,709 cases at the same date. However, I decided not to consider this region in the analysis due to the presence of an outlier on April 4, 2020.
[3] The data are available at URL: https://github.com/pcm-dpc/COVID-19/tree/master/dati-regioni.

## 3. Method and Results

In the last few months an increasing body of literature has attempted to forecast the trend and the final size of the COVID-2019 pandemic by using different approaches (Batista 2020; Benvenuto et al. 2020; Fanelli and Piazza 2020; Giordano et al. 2020; Gupta and Pal 2020; Kumar et al. 2020; Read et al. 2020; Wu et al. 2020; Zhao et al. 2020; Zhou et al. 2020). The autoregressive integrated moving average (ARIMA) model is one of them (Benvenuto et al. 2020; Gupta and Pal 2020; Kumar et al. 2020). It could be considered one of the more used prediction models for time series. In particular, it combines the regressive process and the moving average, and it allows to predict a given time series by considering its own lags and lagged forecast error. The optimal ARIMA model parameters have been chosen i) by using the Akaike's information criterion (AIC); ii) by investigating autocorrelation function (ACF) and partial autocorrelation function (PACF) of the residuals; and iii) by testing the common statistical assumptions about residuals.

Specifically, I follow the approach of He and Tao (2018), Wang et al. (2018), and Benvenuto et al. (2020).[4] In first instance, I check if the Italian regional and national time series are stationary by using the Augmented Dickey-Fuller (1981) (ADF) test and the modified ADF-GLS (or ERS) test for unit root developed by Elliott, Rothenberg and Stock (1992).[5] The tests (table 1) show that almost all variables have a unit root and need to be transformed into a stationary process.

Then, I use Akaike's information criterion (AIC) and the mean absolute error (MAE)[6] to identify ARIMA lag order (p), degree of differencing (d), and order of moving average (q). The best parameters for ARIMA models, according to AIC and MAE minimization, are reported in Table 2. They are the following: Emilia Romagna (0, 2, 1), Marche (0, 2, 2), Lombardy (1, 2, 1), Tuscany (3, 2, 1), Veneto (0, 2, 2), and Italy (4, 2, 2).

Finally, I implement three different tests to perform diagnostic cheeks on the residuals: i) the Doornik and Hansen (1994) test for normality; ii) the Engle's (1982) Lagrange Multiplier test for the ARCH (autoregressive conditional heteroskedasticity) effect; and iii) the Ljung-Box test for autocorrelation. All tests allow to accept the null hypothesis of normality, homoskedasticity, and autocorrelation of the residuals (Table 3).[7]

Table 4 reports the summary of the results of ARIMA models (Figure 2 to 7). The forecast algorithm seems to indicate that the national new daily COVID-2019 cases are largely stabilized and will probably drop near to zero (local cases) in the next 38 days, at least (Figure 2). The hypothetic inflection point[8] will be reach after May 12, 2020, at least. And the final epidemic size should be between 194,000 and 206,000 cases.[9] A similar downward trend is obtained by fitting a specific ARIMA model for the single regions. Specifically, Emilia Romagna requires 42 days to come closer to zero local new cases (Figure 3), Lombardy needs 32 days (Figure 4), Tuscany requires 56 days (Figure 6), and Veneto needs 28 days to significatively flatten the COVID-2019 curve (Figure 7), at least. Marche needs only 12 days (Figure 5), but it does not seem a conservative estimate. So, in Table 4 I report the average mean between the first and the second best ARIMA model. It indicates that Marche needs on average 40 days to reach the hypothetic inflection point. The absence of significant residual spikes in ACF and PACF correlograms, shows that all models are a good fit (Figures A1, A2, and A3 in the Appendix).[10]

---

[4] To carry out the econometric analysis I used Gretl-2020a Software (http://gretl.sourceforge.net/win32/index_it.html).

[5] I use two different approaches because, as stated by Gujarati and Porter (2009), there is no a recognized uniformly powerful test for detecting unit root.

[6] According to Hyndman and Athanasopoulos (2018), MAE is one of the most commonly used scale-dependent measures to assess the forecast accuracy. Moreover, as well as AIC, it's very easy to interpret.

[7] The only exceptions are Emilia Romagna and Lombardy, that are affected by non-normality and autocorrelation, respectively. If the normality is not a necessary condition for forecasting, the violation of the independence assumption may generate some problems.

[8] I mean the inflection point of the cumulative number of COVID-2019 confirmed cases.

[9] The epidemic final size is obtained by summing the original values until April 4, 2020, and the forecast values for the period after April 4, 2020.

However, it's important to stress that this is only a rough guide that requires other updated estimates to be confirmed.

Table 1. Results of ADF and ERS test for unit root.

| Regions (constant + trend) | Daily cases | | At first difference | |
| --- | --- | --- | --- | --- |
| | ADF | ERS | ADF | ERS |
| Emilia Romagna [1] | -1.7823 | -1.9214 | -6.0472*** | -7.4615*** |
| Lombardy [1] | -3.0634 | -2.1693 | -9.2458*** | -9.4513*** |
| Marche [3] | -0.754 | -1.0985 | -7.091*** | -10.3264*** |
| Tuscany [1] | -4.5378*** | -4.4087*** | -6.9036*** | -9.3454*** |
| Veneto [2] | -1.473 | -1.7598 | -11.6103*** | -11.8692*** |
| Italy [1] | -1.547 | -1.6586 | -5.8425*** | -5.9628*** |

Notes: lags in brackets. For lag length selection I used AIC approach. Significance level: 0.01***; 0.05**; 0.1*.

Table 2. The optimal parameters for ARIMA models.

| Regions | AR-I-MA parameters | AIC value | Mean absolute value |
| --- | --- | --- | --- |
| Emilia Romagna | (0, 2, 1) | 476.9949 | 64.995 |
| Lombardy | (1, 2, 1) | 642.7858 | 306.25 |
| Marche | (0, 2, 2) | 380.2007 | 27.491 |
| Tuscany | (3, 2, 1) | 428.1028 | 38.716 |
| Veneto | (0, 2, 2) | 477.2967 | 60.88 |
| Italy | (4, 2, 2) | 650.9967 | 334 |

Notes: for parameter selection I used AIC approach.

---

[10] The only exception is Lombardy and Veneto, that have two (lag 4 and 11) and one (lag 3) significant spikes, respectively.

Table 3. The results of normality, ARCH, and autocorrelation tests for the ARIMA models (Figures 2-7).

| Regions | Doornik-Hansen test for normality | | Engle's LM test for ARCH effect | | Ljung Box test for autocorrelation | |
|---|---|---|---|---|---|---|
| | Value | p-value | Value | p-value | Value | p-value |
| Emilia Romagna | 12.467 | 0.002 | 0.8611 | 0.973 | 13.1893 | 0.1542 |
| Lombardy | 0.966 | 0.6168 | 14.9047 | 0.1356 | 14.783 | 0.0388 |
| Marche | 4.148 | 0.1257 | 5.2826 | 0.8715 | 5.6599 | 0.6853 |
| Tuscany | 1.645 | 0.4393 | 5.7887 | 0.8327 | 5.9615 | 0.4275 |
| Veneto | 2.103 | 0.3493 | 3.4188 | 0.9698 | 13.0181 | 0.1112 |
| Italy | 2.775 | 0.2496 | 5.008 | 0.8906 | 4.9098 | 0.2967 |

Notes: for lag selection I followed Hyndman and Athanasopoulos (2018), that suggest a value of 10.

Table 4. Summary of the results of ARIMA models (Figure 2-7).

| Regions | Inflection point (days since April 4) | Inflection point (date) | Epidemic final size (approximate) |
|---|---|---|---|
| Emilia Romagna | 42 | May 16, 2020 | 32,600 to 33,400 |
| Marche (mean)* | 40 | April 16, 2020 | 6,300 to 7,100 |
| Lombardy | 32 | May 6, 2020 | 77,000 to 80,000 |
| Tuscany | 56 | May 30, 2020 | 14,400 to 15,200 |
| Veneto | 28 | May 2, 2020 | 15,300 to 16,000 |
| Italy | 38 | May 12, 2020 | 194,000 to 206,000 |

Notes: *this is the average mean between the first and the second best ARIMA model for Marche.

Figure 2. Results of ARIMA forecast approach for overall national data.

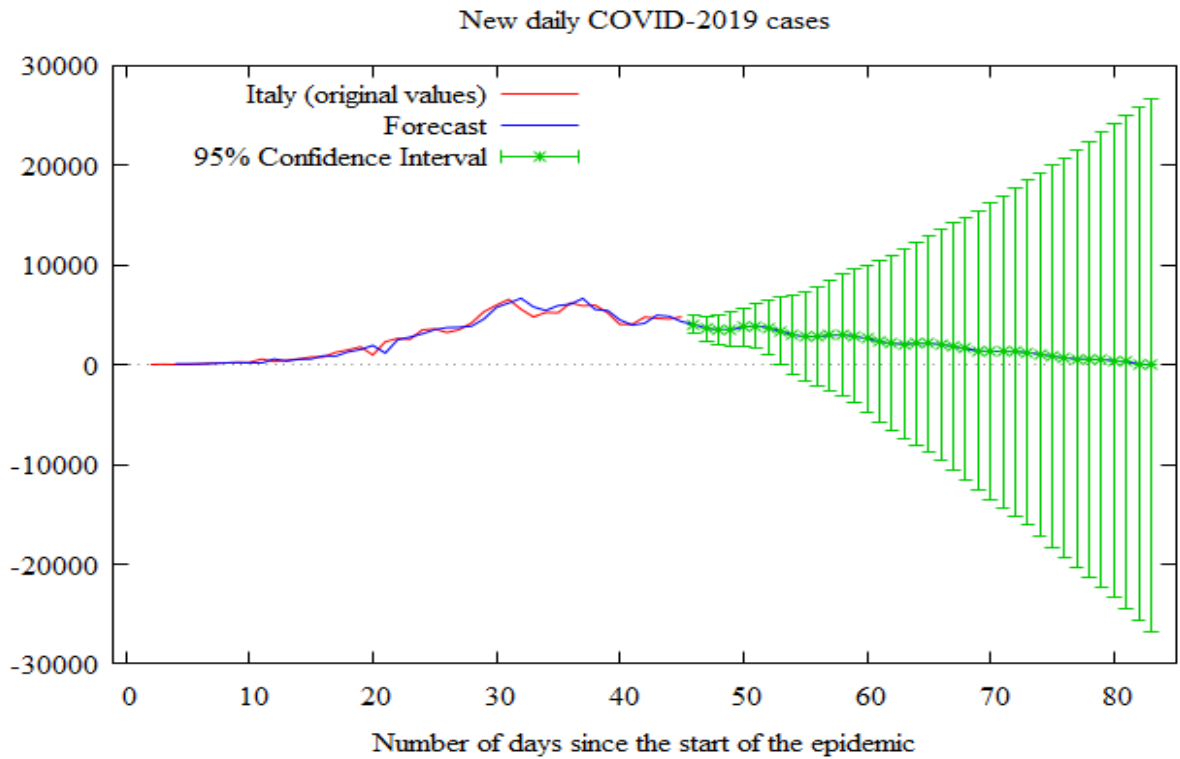

Figure 3. Results of ARIMA forecast approach for Emilia Romagna.

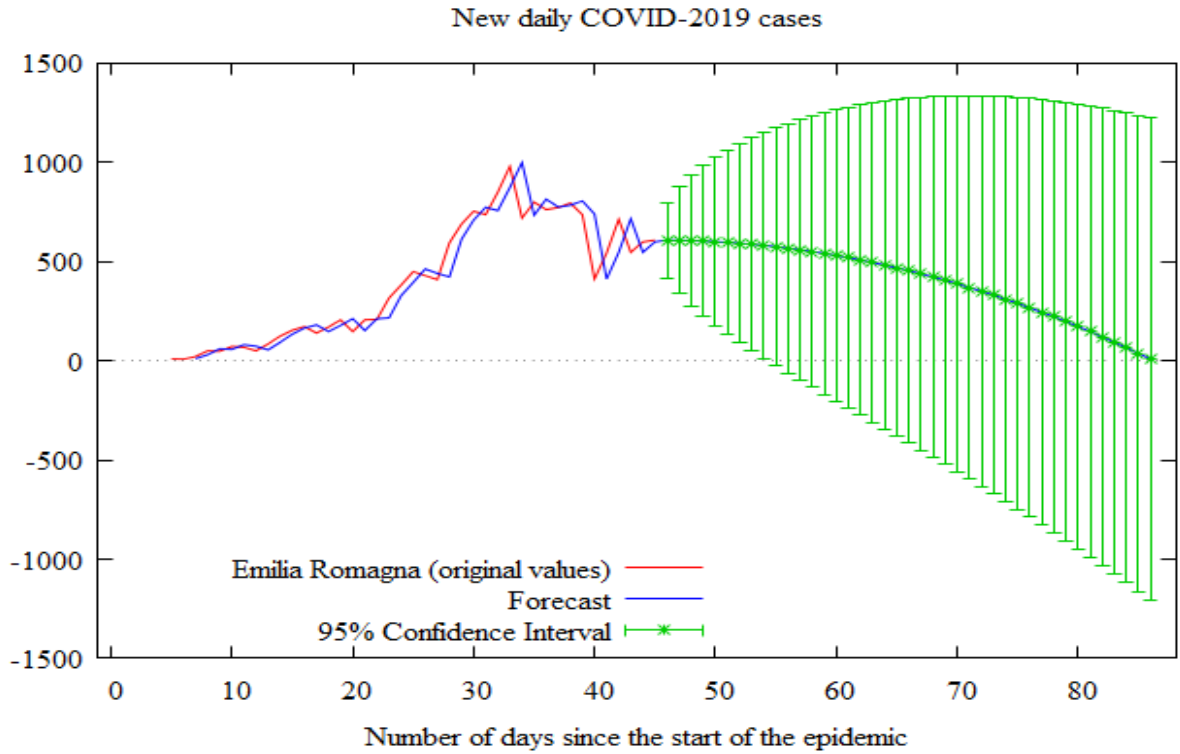

Figure 4. Results of ARIMA forecast approach for Lombardy.

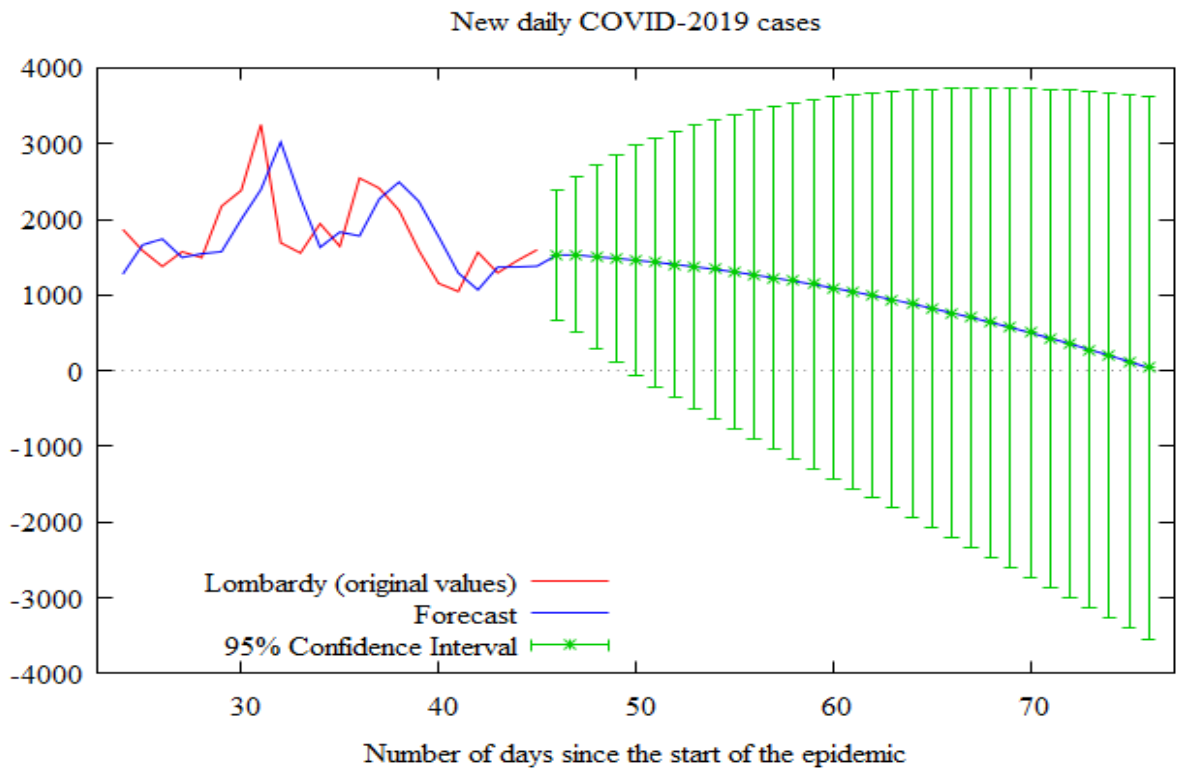

Figure 5. Results of ARIMA forecast approach for Marche.

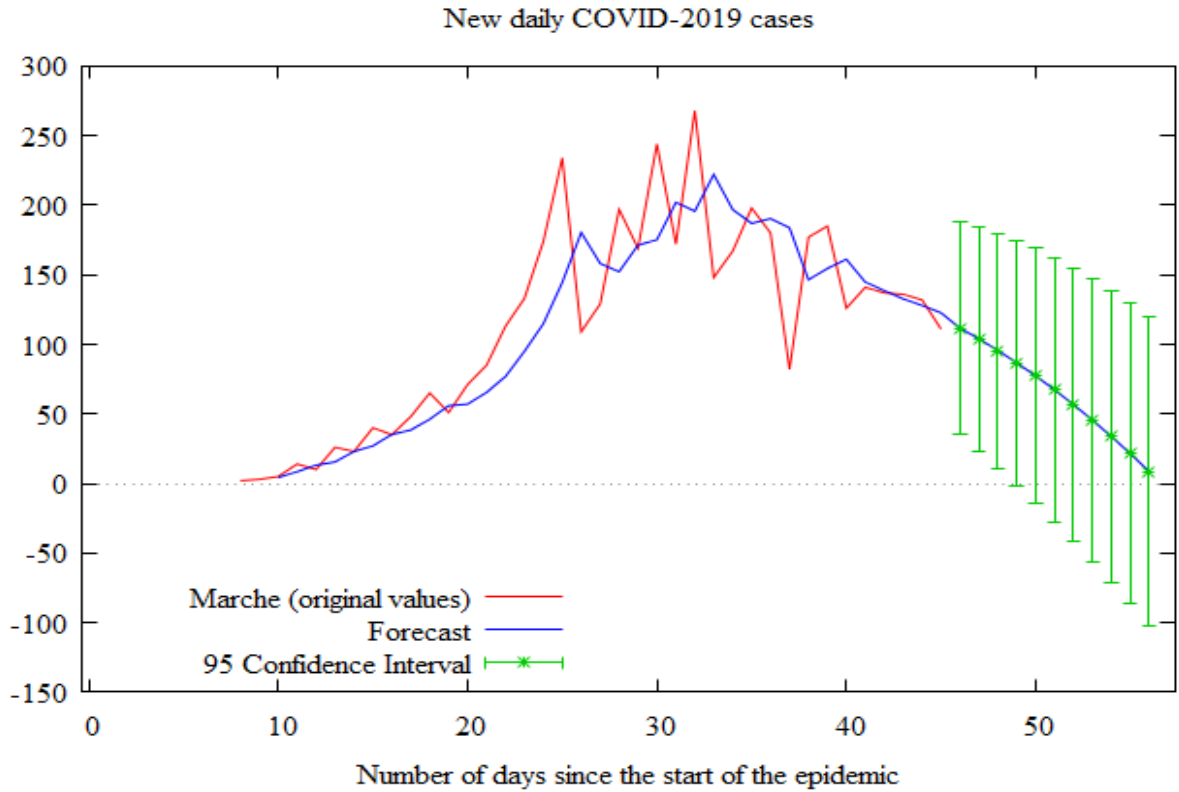

Figure 6. Results of ARIMA forecast approach for Tuscany.

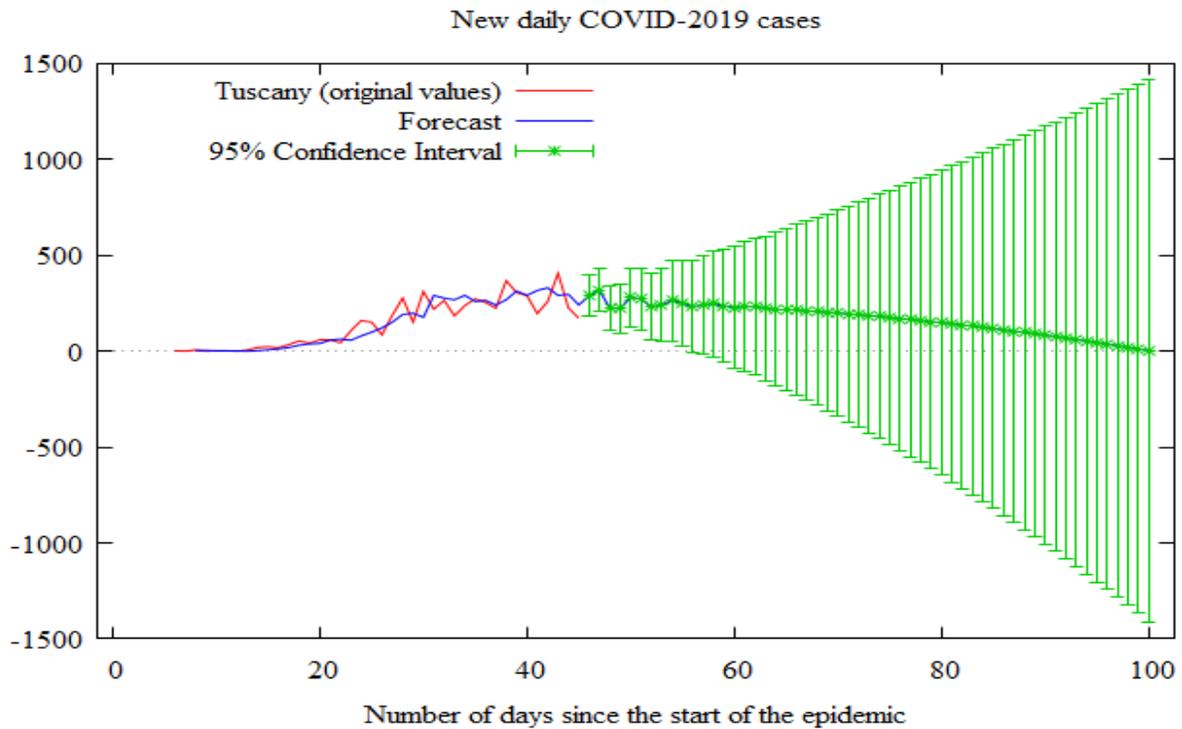

Figure 7. Results of ARIMA forecast approach for Veneto.

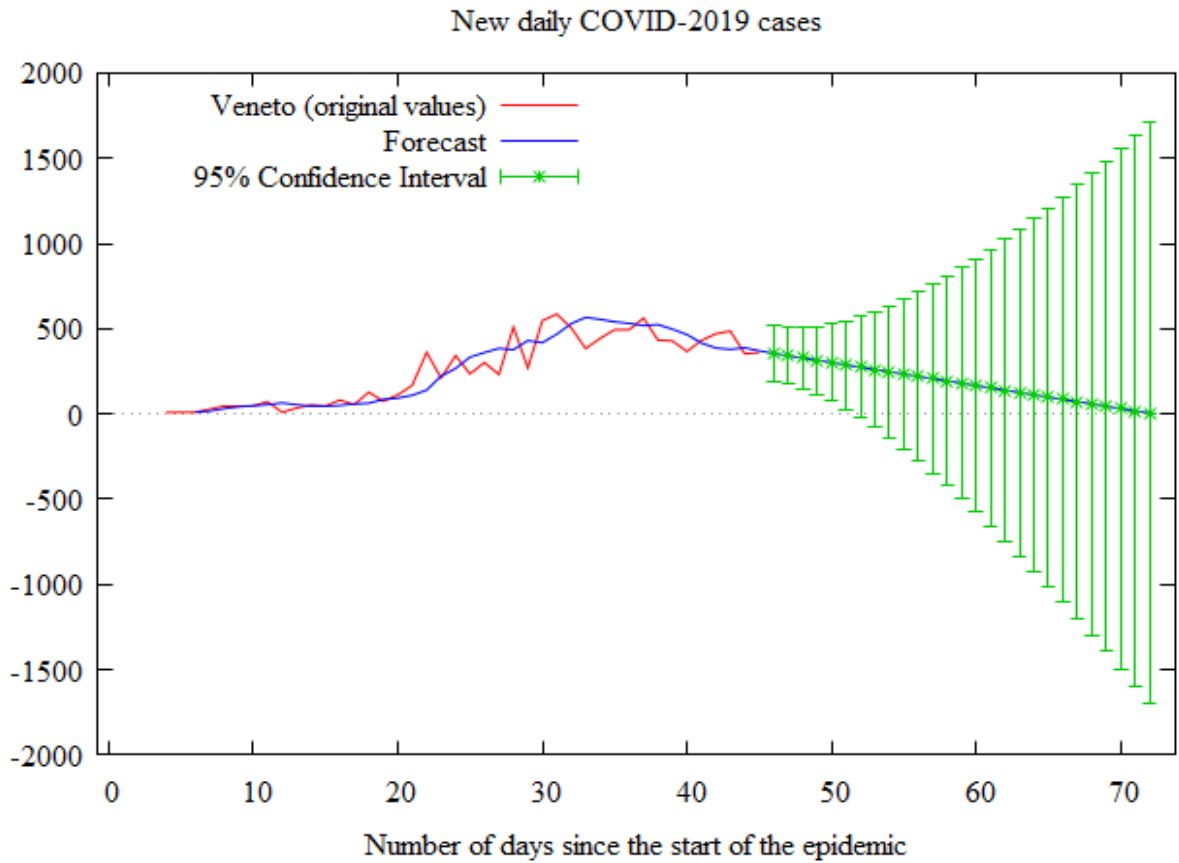

## 4. Conclusion

ARIMA models can be viewed as a simple and immediate tool to program the health monitoring system at national and regional level. The main advantages of ARIMA forecasting approach are surely its ease of application and interpretation. By the contrary, it is sensitive to outliers in the data and, do not account for the noise, that is unknown by definition. For these reasons it may be considered a good model for short-term forecasting, but the results should be interpreted with thriftiness.

Results suggest that COVID-2019 epidemic in Italy will reach the *plateau*, in term of cumulative cases, in the next 40-55 days, i.e. about the entire month of April and May 2020. Specifically, Lombardy and Veneto seem to require a lower number of days than the other regions, especially compared to Tuscany, that will need approximately 56 days to definitively flatten the COVID-2019 curve. The final epidemic size in Italy should be around 200,000 cases. However, it is necessary to stress that this rough estimation is strongly related to the previous values. The continuation of the restrictive measures and the strict compliance with the rules, such as traffic and travel restriction, ban on gatherings, and closure of commercial activities, may mitigate the size of the epidemic.

Further useful and more precise forecasting may be provided by updating these data and applying the model to other regions and countries.

# Appendix

Figure A1. ACF and PACF correlograms for Italy (on the left) and Emilia Romagna (on the right).

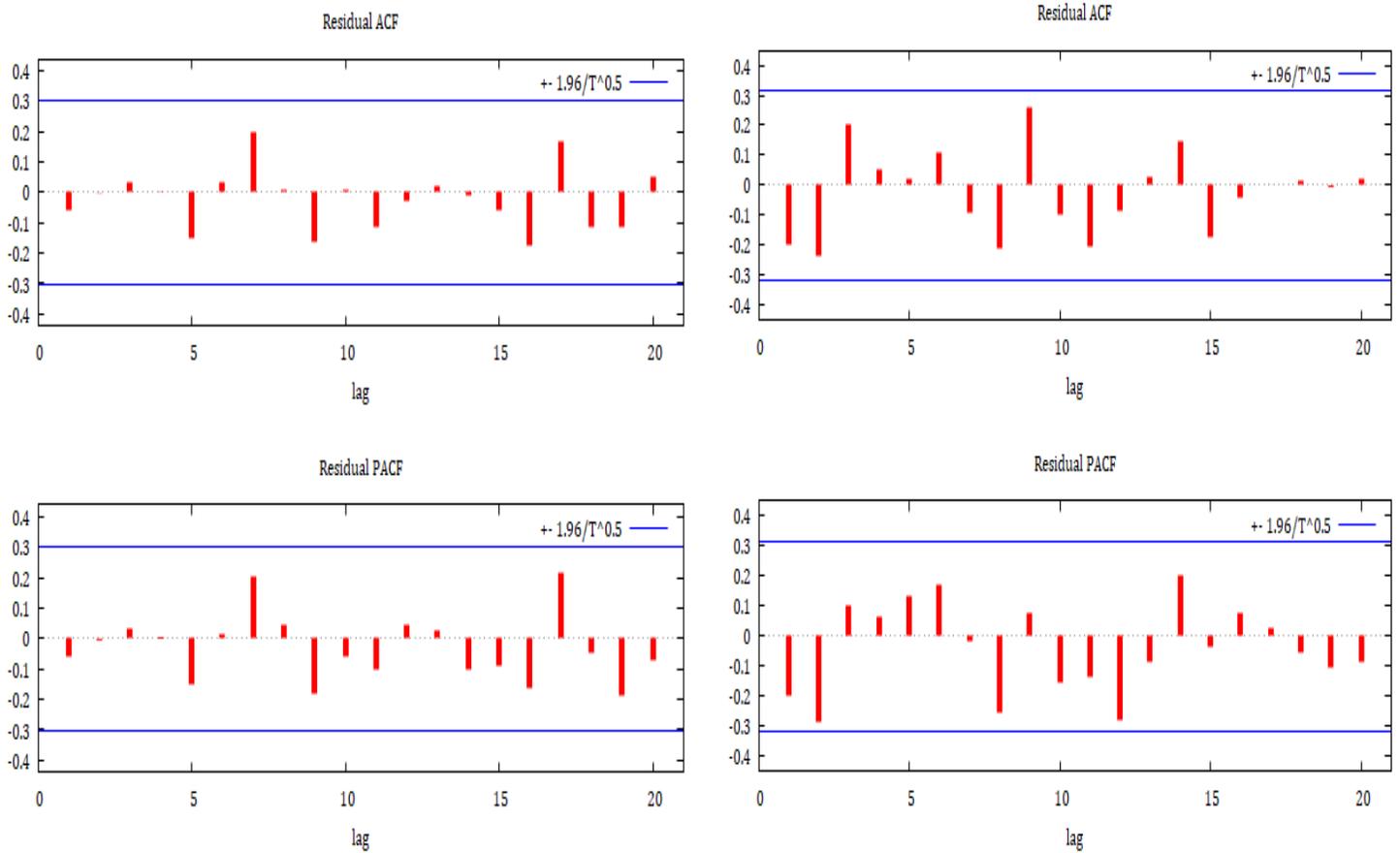

Figure A2. ACF and PACF correlograms for Lombardy (on the left) and Marche (on the right).

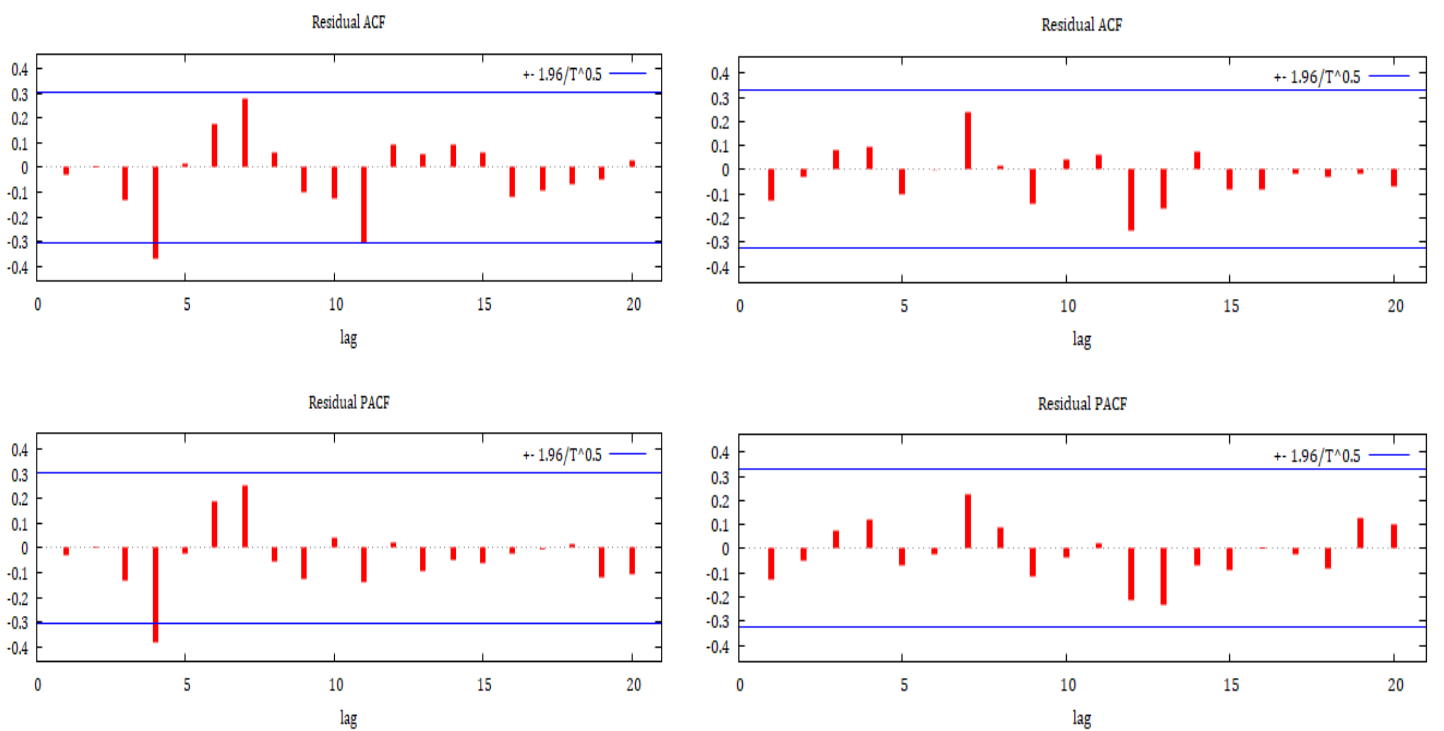

Figure A3. ACF and PACF correlograms for Tuscany (on the left) and Veneto (on the right).

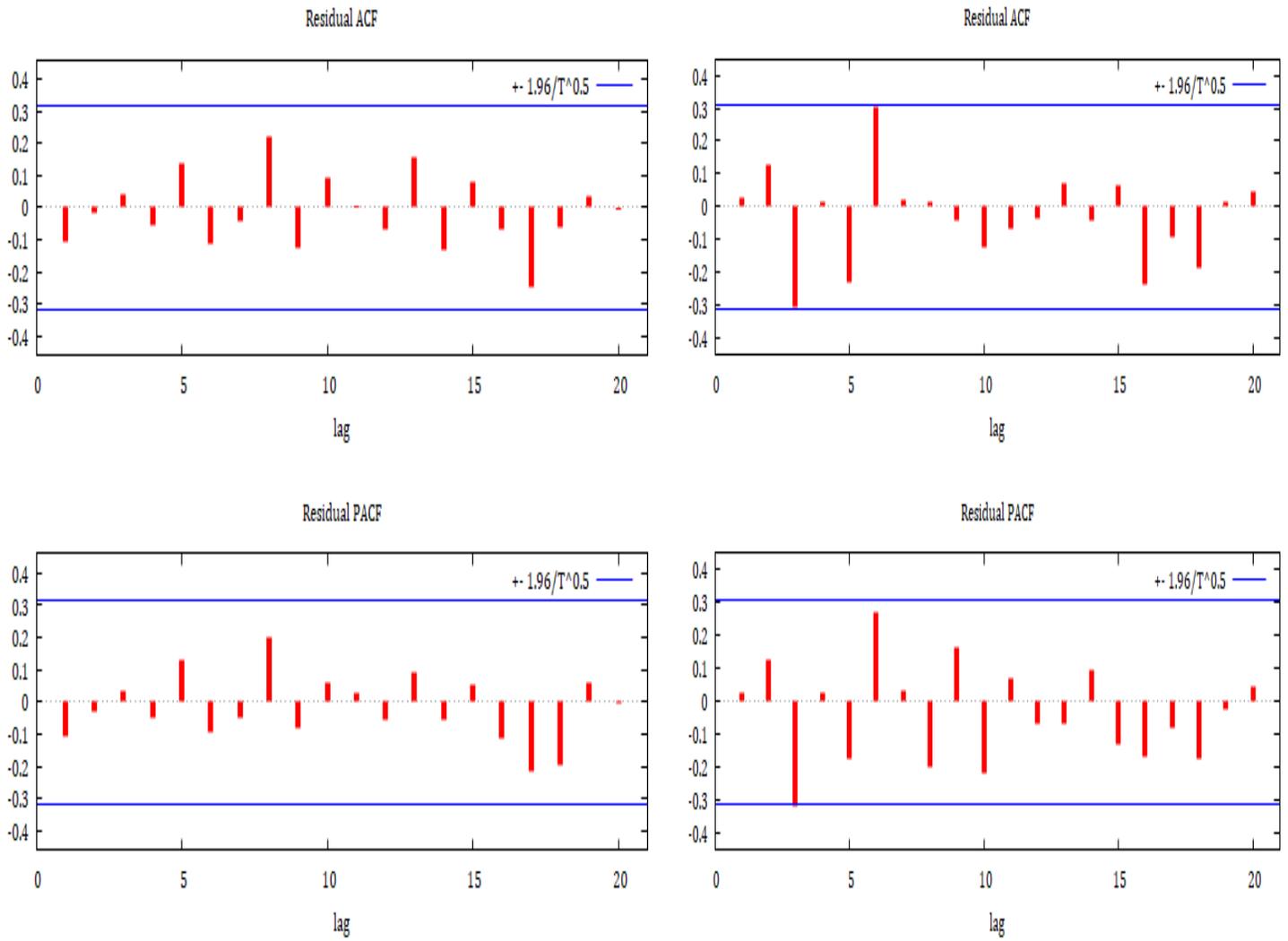